# Investigating Issues that Lead to Code Technical Debt in Machine Learning Systems

Rodrigo Ximenes
*Department of Informatics*
PUC-Rio
Rio de Janeiro, Brazil
rximenes@inf.puc-rio.br

Antonio Pedro Santos Alves
*Department of Informatics*
PUC-Rio
Rio de Janeiro, Brazil
apsalves@inf.puc-rio.br

Tatiana Escovedo
*Department of Informatics*
PUC-Rio
Rio de Janeiro, Brazil
tatiana@inf.puc-rio.br

Rodrigo Spinola
*Department of Computer Science*
Virginia Commonwealth University (VCU)
Richmond, United States of America
spinolaro@vcu.edu

Marcos Kalinowski
*Department of Informatics*
PUC-Rio
Rio de Janeiro, Brazil
kalinowski@inf.puc-rio.br

*Abstract*—[Context] Technical debt (TD) in machine learning (ML) systems, much like its counterpart in software engineering (SE), holds the potential to lead to future rework, posing risks to productivity, quality, and team morale. Despite growing attention to TD in SE, the understanding of ML-specific code-related TD remains underexplored. [Objective] This paper aims to identify and discuss the relevance of code-related issues that lead to TD in ML code throughout the ML workflow. [Method] The study first compiled a list of 34 potential issues contributing to TD in ML code by examining the phases of the ML workflow, their typical associated activities, and problem types. This list was refined through two focus group sessions involving nine experienced ML professionals, where each issue was assessed based on its occurrence contributing to TD in ML code and its relevance. [Results] The list of issues contributing to TD in the source code of ML systems was refined from 34 to 30, with 24 of these issues considered highly relevant. The *data pre-processing* phase was the most critical, with 14 issues considered highly relevant. Shortcuts in code related to typical pre-processing tasks (*e.g.*, handling missing values, outliers, inconsistencies, scaling, rebalancing, and feature selection) often result in "patch fixes" rather than sustainable solutions, leading to the accumulation of TD and increasing maintenance costs. Relevant issues were also found in the *data collection*, *model creation and training*, and *model evaluation* phases. [Conclusion] We have made the final list of issues available to the community and believe it will help raise awareness about issues that need to be addressed throughout the ML workflow to reduce TD and improve the maintainability of ML code.

## I. Introduction

The concept of technical debt (TD), first introduced by Ward Cunningham in 1992 [7], uses an analogy between software engineering (SE) and financial borrowing. This analogy highlights how opting for quicker deliveries in SE can lead to accumulated 'debt' that, similar to its financial counterpart, accrues 'interest' over time and demands eventual repayment. Leveraging TD can accelerate project timelines, especially when focusing on developing features rapidly. However, rigorous management and timely repayment are essential to minimize the risks associated with these expedited processes.

The intersection of TD and machine learning (ML) systems has emerged as a noteworthy area of interest, as emphasized by Sculley *et al.* [20]. Despite this growing attention, understanding and managing TD within the context of ML systems remain relatively unexplored, making it an area in need of further investigation. Martinez *et al.* [12] identified several key challenges in executing real-world ML projects, including low process maturity, reproducibility issues, absence of validation data, and inadequate quality assurance checks. Overcoming these challenges underscores the importance of cohesive coordination among team and project management and efficient data information management to ensure reproducibility, reliability, and the successful realization of project objectives. In another study, Pimentel *et al.* [16] uncovered sub-optimal practices in the use of notebooks, which are widely adopted in ML-enabled system projects for coding [10].

Motivated by these gaps, this study explores potential causes of ML code-related TD accumulation, aiming to raise aware- ness among ML developers about the risks of accumulating code-related TD during typical tasks throughout the ML development workflow. Specifically, we focus on answering the following three research questions (RQs): *RQ1 - What are potential issues that could lead to ML code TD?*; *RQ2 - To what degree do the identified issues manifest in practice, con- tributing to code-related TD?*; and *RQ3 - How do practitioners perceive the relevance of the issues in terms of leading to TD?* To answer these research questions, we examined typical activities of each phase of the ML workflow [2] and potential problem types [8], identifying 34 candidate issues that could contribute to ML code TD. These issues were then evaluated for their occurrence in practice leading to TD and relevance through two focus group sessions involving nine experienced ML practitioners.

Insights from the focus group sessions refined the list to 30 issues considered to lead to ML code TD, with 24 deemed highly relevant. We found that TD in ML code arises from issues across different phases of the ML workflow. In the *data collection* phase, improper integration and insufficient data validation often lead to incomplete or biased datasets, resulting in unreliable models and costly rework. The *data pre-processing* phase is particularly critical, with issues in handling missing values, outliers, inconsistencies, scaling, rebalancing, and feature selection propagating through the system, often leading to temporary workarounds rather than sustainable solutions. During the *model creation and training* phase, inadequate splitting of datasets, neglected hyperparameter tuning, and insufficient algorithm testing can lead to suboptimal models that require repeated adjustments. Lastly, in the *model evaluation phase*, the misalignment of evaluation metrics with project goals can result in misleading assessments of model performance, requiring substantial re-engineering efforts.

The remainder of this paper is organized as follows. In Section II, we provide the background and an overview of relevant prior research. In Section III, we present an overview of the research method. In Section IV, we outline the compilation of the candidate issues. In Section V we detail the planning of the focus group assessments. In Section VI, we detail the assessment and refinement of the list of candidate issues and summarize the results. Subsequently, we discuss the results in Section VII and threats to validity in VIII. Finally, in Section IX, we present our concluding remarks and potential avenues for future research.

## II. BACKGROUND AND RELATED WORK

### A. Technical Debt in Machine Learning Systems

In software development, Ward Cunningham [7] introduced the concept of TD, likening it to financial debt that allows for faster initial development but incurs "interest" to be paid later. Accumulating TD has far-reaching consequences. According to Tom *et al.* [22], TD reduces productivity by making maintenance and code modifications more challenging. It can negatively impact developers' morale, as they need to invest extra effort in dealing with it. Over time, developers must repay the "interest" and "principal" on the accrued debt to maintain their system. In a systematic literature review, Alves *et al.* [1] gathered various types of TD. The primary focus of this work is on Code TD, which concerns issues within the source code that hinder maintenance and may lead to additional effort in the future.

In the realm of artificial intelligence (AI) and ML, Bogner *et al.* [5] conducted a systematic mapping study based on 21 primary studies and identified four new types of TD that emerge in AI-based systems: data debt, model debt, configuration debt, and ethics debt. They also noted that conventional TD types like infrastructure, architectural, code, and test debt also apply to AI systems. However, they may have AI-specific aspects, such as managing and monitoring AI pipelines and models to mitigate infrastructure debt. Recupito *et al.* [17] analyzed technical debt in AI-enabled systems, highlighting the prevalence and impact of code and architectural debt, as well as management strategies used to address these challenges.

Tang *et al.* [21] recognized a knowledge gap regarding the evolution and maintenance of ML systems. They conducted a study across 26 projects, identifying seven new TD categories specific to ML: custom data types, duplicated feature extraction code, model code reusability, unnecessary model code, model code comprehension, model code modifiability, and duplicated model code. They found that duplication significantly contributes to TD in ML systems, particularly in configuration and model code. Furthermore, Zhang and Cruz [23] explored code smells specific to ML applications, emphasizing how they manifest in ML pipelines and contribute to technical debt, particularly in terms of maintainability and reusability.

Tang *et al.* also highlight that while traditional software developers frequently use inheritance to reduce code duplication, the increasing popularity of scripting languages, mostly used in ML systems, requires model code to be written in an object-oriented manner. This transition poses challenges for ML developers in incorporating inheritance effectively. In line with this, Cabral *et al.* [6] provided evidence that the adoption of object-oriented design principles can improve code understanding within ML projects, also enhancing the maintainability of ML code.

While previous work uncovered new types of technical debt for ML-enabled systems, it did not explore issues related to the intricacies of typical ML workflow activities that may lead to ML code TD, which is the focus of this work.

### B. Machine Learning Workflow

Machine Learning (ML) workflows are systematic processes that transform raw data into trained models capable of making informed predictions or decisions. These workflows are key to numerous applications, and a typical ML workflow consists of several key phases. Amershi et al. [2] present a nine-phase machine learning workflow, including requirements, data collection, data cleaning, data labeling, feature engineering, model creation and training, model evaluation, model deployment, and model monitoring.

We used this nine-phase ML workflow as a basis to ground our discussions but, while remaining consistent with it, as done by Kalinowski *et al.* [9], we abstracted *data cleaning*, *data labeling*, and *feature engineering* into a more generic *data pre-processing* phase. The rationale was that these activities are typically part of pre-processing and that this abstraction allows flexibility to discuss different pre-processing strategies. For instance, unsupervised ML algorithms do not require data labeling. Hence, we adopted the seven more generic ML phases to characterize typical machine learning workflows shown in Figure 1.

In summary, after *problem understanding and requirements*, the ML workflow follows with *data collection* and *data pre-processing*, where data is gathered, cleaned, and transformed, with feature engineering enhancing data relevance. Following this, *model creation and training* starts with selecting the most suitable ML algorithm and optimizing its hyperparameters to allow the model to learn patterns in the data. Thereafter, *model evaluation* is performed on a validated dataset by analyzing various metrics. After training and evaluation, *model deployment* is responsible for serving the model in production, and *model monitoring* is responsible for continuous monitoring to detect performance degradation due to phenomena like data drift, ensuring sustained reliability.

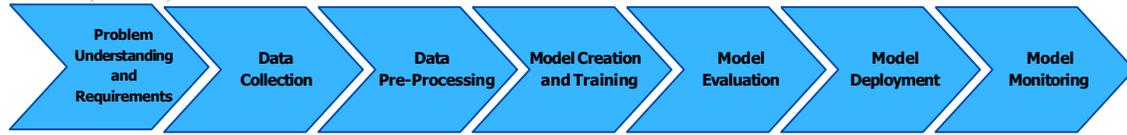

Fig. 1: Seven typical phases of the ML workflow

## III. RESEARCH METHOD

### A. Goal and Research Questions

Our main research goal was to compile and assess a list of issues that can lead to ML code TD. Therefore, our first research question concerns compiling the list.

*RQ1: What are potential issues that could lead to ML code TD?* This research question aims to derive a set of potential issues that can lead to ML code TD by analyzing problems related to typical activities conducted throughout the different ML workflow phases.

After compiling such a list, we wanted to assess the occurrence and relevance of the identified issues from the point of view of ML experts. Therefrom, we derive the following two additional research questions:

*RQ2: To what degree do the identified issues manifest in practice, contributing to code-related TD?* This research question aims to verify whether the proposed issues manifest in real-world ML systems and if they can lead to code-related TD.

*RQ3. How do practitioners perceive the relevance of the issues in terms of leading to TD?* This research question evaluates the perceived relevance of the suggested issues in the candidate list regarding leading to TD.

To assess these RQs from a practitioner's point of view, we conducted two focus group sessions with nine ML experts.

### B. Research Method Overview

To address the research questions, we defined the following two-step research method:

*Compilation of the Candidate Issues.* The first author analyzed potential problems related to typical coding activities conducted throughout the different ML workflow phases to build a list of issues that may lead to code-related TD. The list was refined in discussions with an independent researcher. Subsequently, this list was reviewed by an additional independent researcher to eliminate inconsistencies and add any overlooked issues. Further details on the compilation and the candidate list can be found in Section IV.

*Focus Group Sessions.* We conducted two focus group sessions, following the guidelines by Kontio *et al.* [11] to facilitate in-depth discussions and obtain insights from ML experts, with the objective of assessing the occurrence and relevance of the candidate issues. The focus group sessions were conducted with nine ML experts. Further details on the focus group sessions and their results can be found in Sections V and VI.

## IV. COMPILATION OF THE CANDIDATE ISSUES

Practitioners conduct various activities across the ML workflow that result in ML code. For instance, during the *pre-processing phase*, typical activities include identifying and handling outliers and missing values, feature selection, rebalancing, removing inconsistent data, and pre-processing scaling (*e.g.*, normalization, standardization). Throughout *model creation and training*, practitioners conduct activities including selecting candidate ML algorithms, hyperparameter tuning, and cross-validation to ensure adequate performance.

Gathering these common activities across different ML workflow phases, Table I was conceived. Even though not all activities are executed in every ML project, they reflect typical activities ML practitioners conduct throughout the ML workflow. This table is structured with five columns: the first denotes the specific phase in the ML workflow, the second column describes a typical activity, and the other three columns represent categories of problems (*Missing*, *Incomplete/Insufficient*, and *Inappropriate/Wrong*) taken from the IEEE Standard for Software Anomalies [8].

The combination of the activity with the categories of problems generates our initial list of candidate issues. For instance, *Missing identifying outliers* refers to not identifying outliers, *Incomplete/Insufficient identifying outliers* refers to not investing sufficient effort in this activity, or *Inappropriate/Wrong identifying outliers* refers to applying inappropriate or wrong methods for identifying outliers. Note that each of these example issues will be reflected in the ML code responsible for data pre-processing. This suboptimal pre-processing might still allow an ML model to be created and deployed but can cause added effort in the evolution of the ML-enabled system over time (*e.g.*, in the face of performance degradation, the sub-optimal pre-processing might have to be revisited and improved).

Note that this crossing of activities with categories of problems resulted in some combinations that do not make sense in real-world (*e.g.*, *Incomplete/Insufficient Rebalancing* - either the practitioner did not rebalance, or he did it wrong), so they were not considered. These are marked with an '✗' in Table I, while all possible combinations are marked with an '✓.' Furthermore, no ML code-related issues were identified for the *Problem Understanding and Requirements*, *Model Deployment*, and *Model Monitoring* ML workflow phases, which typically do not involve activities that directly produce ML code.

The final list contained 34 candidate issues to be presented to the experts and discuss their occurrence in practice contributing to code-related TD and their relevance in terms of leading to such TD.

## V. Focus Group Planning

### A. Focus Group Main Goal and Scope

The main goal of the focus group sessions is to confirm if the candidate issues occur in practice and if they are perceived as relevant by practitioners in terms of leading to TD. The GQM (Goal Question Metric) definition template [4] is used to set this goal as follows: **Analyze the candidate list of issues that can lead to ML code TD with the purpose of characterizing with respect to the occurrence in practice contributing to TD and perceived relevance of the issues from the point of view of ML experts in the context of developing ML systems**.

### B. Focus Group Population

We employed a targeted convenience sampling approach for our population of ML experts, as we had direct access to these professionals. We extended invitations to individuals we knew possessed relevant experience and were actively involved in developing ML-enabled systems. Specifically, we selected participants who were actively engaged in ML projects at ExACTa PUC-Rio, Petrobras, and Serpro. ExACTa PUC-Rio concerns an R&D laboratory with approximately 100 collaborators working with several industry partners. Petrobras is a corporation that operates in the energy industry and has approximately 40,000 employees. Serpro is a public company providing information technology services to the government of Brazil and currently has approximately 10,000 employees.

### C. Focus Group Preparation

We used the online tool Miro[1] to create a virtual interactive focus group discussion board, as depicted in Figure 2, to streamline the orchestration of the focus group session.

The board comprised four columns. The first indicates the phase of the ML workflow. The other three columns represent categories of problems (e.g., Missing, Incomplete/Insufficient, and Inappropriate/Wrong), as described in Section IV. Each line represented one activity and each cell represented one issue to be discussed, alongside two questions concerning its occurrence and relevance and space for post-it comments.

We employed the Likert scale (1- Disagree, 2- Partially Disagree, 3- Not Sure, 4- Partially Agree, and 5- Agree) to assess participants' levels of agreement with the two questions posed for all candidate issues. The first question was **It may occur in practice contributing to TD**, indicating the likelihood of the issue leading to code-related TD. The second question was **I consider it relevant**, indicating the relevance of TD resulting from the issue, assuming it occurs.

[1]https://miro.com/

Each participant was represented by a colored dot, which allowed them to move freely to their preferred answer on the Likert scale. Additionally, a green box containing explanations of the activity was provided on the right-hand side for clarity and to eliminate any uncertainties regarding the activity. Moreover, participants could comment by adding sticky notes to each proposed TD item. This interactive approach facilitated the discussions during the focus group sessions.

### D. Focus Group Session Dynamics

Following the recommendation of Menary *et al.* [14] for online focus-group sessions, we organized multiple sessions with limited participants. To accommodate this, we divided the participants into two focus group sessions based on availability and preferences.

The two remote focus group sessions took place via video conferences. The video conferences were held using the Zoom platform and recorded to facilitate getting all relevant participants' comments. Both sessions had approximately 2 hours in duration.

These sessions included nine industry practitioners, with four participants in the first and five in the second. Additionally, two researchers (the first and the last author) served as facilitators throughout both sessions. The session dynamics reflected following these steps:

1) A participant characterization form was distributed and filled;
2) A brief presentation was included to provide participants with a clear definition of TD;
3) Subsequently, the facilitators introduced the Miro board and explained how the session would be conducted;
4) Finally, the facilitators proceeded to individually introduce each issue for the experts to discuss and register their opinions and comments.

## VI. Focus Group Results

### A. Participant Characterization

Table II shows the participants' characterization. It is possible to observe that the recruited participants are experienced data scientists with high educational levels. The first focus group session included four participants, P1 to P4, while the second involved five participants, P5 to P9.

### B. Assessing the Issues for each ML Workflow Phase

We present the assessment of the issues following a structured approach: for each ML workflow phase, we group the discussion and outcomes for each candidate issue per activity. During these discussions, we visually exhibit the final votes cast in each session for each issue and incorporate expert comments made during the sessions to enhance comprehension of why a certain issue was considered to occur leading to ML code TD (or not) and why it was considered relevant (or not).

We used the median value to determine the occurrence and relevance of each issue. For occurrence, we considered that the issue occurs leading to TD if the median was between *partially agree* and *agree*. Regarding relevance, we considered the relevance high if the median was *agree*; otherwise, we considered it low.

TABLE I: Candidate issues that may lead to ML code TD

| Phase | Activity | Missing | Incomplete/Insufficient | Inappropriate/Wrong |
|---|---|---|---|---|
| Data collection | Data integration | ✓ | ✓ | ✓ |
|  | Data consumption | ✓ | ✓ | ✓ |
| Data pre-processing | Identifying outliers | ✓ | ✓ | ✓ |
|  | Selecting features when needed | ✓ | ✓ | ✓ |
|  | Identifying missing values | ✓ | ✓ | ✓ |
|  | Rebalancing (typically by resampling) | ✓ | ✗ | ✓ |
|  | Removing inconsistent data (remove the inconsistency) | ✓ | ✓ | ✓ |
|  | Pre-processing scaling | ✓ | ✓ | ✓ |
| Model creation and training | Testing candidate algorithm possibilities | ✓ | ✓ | ✓ |
|  | Splitting training/test/validation data | ✓ | ✗ | ✓ |
|  | Hyperparameter tuning | ✓ | ✓ | ✓ |
| Model evaluation | Choosing evaluation metric | ✗ | ✓ | ✓ |
|  | Properly using methods for evaluating a model's performance | ✗ | ✗ | ✓ |

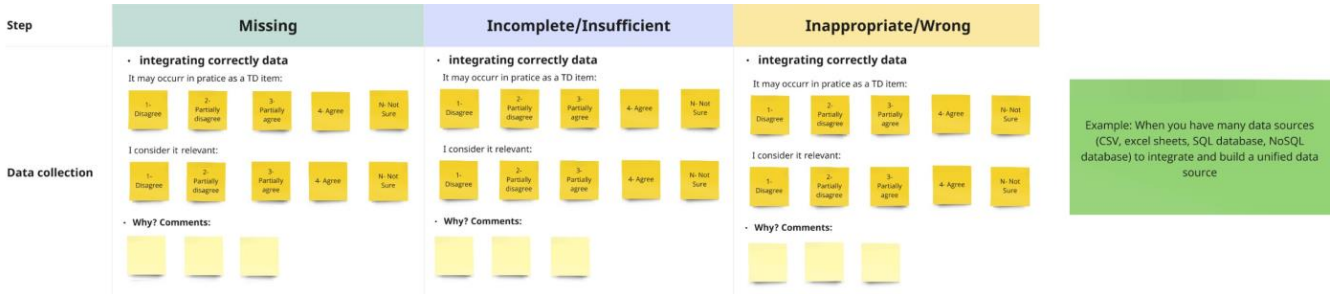

Fig. 2: Miro discussion template for the candidate issues

TABLE II: Participant characterization

|  | 1st Session | | | | 2nd Session | | | | |
|---|---|---|---|---|---|---|---|---|---|
| Participant | P1 | P2 | P3 | P4 | P5 | P6 | P7 | P8 | P9 |
| Education level | PhD | MSc | MSc | MSc | PhD | MSc | BSc | BSc | BSc |
| Years of experience with ML systems | 14 | 3 | 3 | 4 | 10 | 4 | 4 | 1 | 2 |
| Number of ML projects | 15 | 10 | 3 | 5 | 15 | 1 | 3 | 3 | 1 |

It is also important to mention that the final votes and post-it comments of the participants for each session can be found in the filled Miro board available in our open science repository [3]. Here, to be concise, we only present the aggregated results considering both sessions. Furthermore, due to space constraints and to communicate our results more objectively, we present a detailed assessment and refinement analysis only for the issues of the Data Collection ML workflow phase. Similar analyses for the issues of all other ML workflow phases can be found in our Open Science Repository [3].

### C. Assessing the Issues for Data Collection ML Phase

**Activity: Data Integration**. This activity concerns integrating data from multiple data sources (*e.g.*, CSV, Excel sheets, SQL databases, NoSQL databases).

1) *Missing data integration:*
   Figure 3 illustrates the aggregated votes of the experts for this issue. It can be observed that its occurrence in practice leading to TD was confirmed. Its relevance was acknowledged to varying degrees, but using the median and following our criteria, it can be classified as an issue of low relevance.

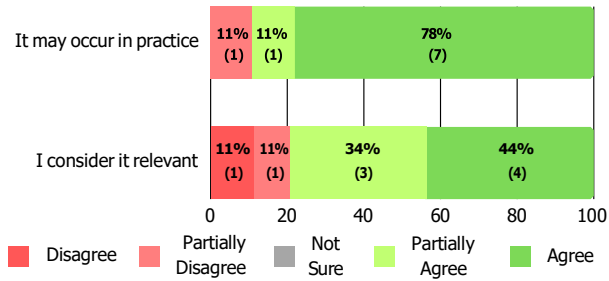

Fig. 3: Missing data integration

Participant P5 gave an example: *"In the context of predictive models, a lack of data integration can lead to the absence of relevant information for the model, hampering the achievement of high accuracy. For example, in practice, a crucial explanatory column may be missing, undermining the model's ability to adequately capture relationships between the available variables"*. Participant P8 reaffirmed the previous comment: *"There will be a possibility of future rework because, when evaluating the results, it may be challenging to explain or find a justification for the inadequate model due to the absence of a portion of the data"*.

2) *Incomplete/Insufficient data integration:*

Figure 4 provides the aggregated votes of the experts for this issue, leaving no doubt about its status as a highly relevant issue.

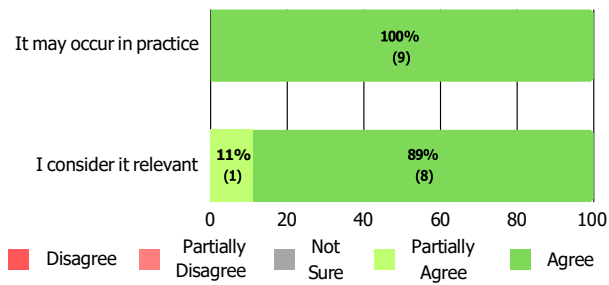

Fig. 4: Incomplete/Insufficient data integration

Participant P2 mentioned a previous experience: *"This reminded me of a situation I went through. There were two data sources. We could access one data source but often lost access to the other. To illustrate what this integration could be, we had to generate different models because we were working with only a portion of the data"*.

Participant P7 argued: *"I think it's quite common and happens more often than simply missing (the previous issue). You consider one part and later realize that something is missing. I believe it's more common than the first scenario because when you do nothing, you immediately see that a step in the process is missing."*.

3) *Inappropriate/Wrong data integration:*

Upon analyzing the results from both sessions, Figure 5 unmistakably establishes the classification of this issue as highly relevant. In both sessions, all participants reported agreement in response to both questions.

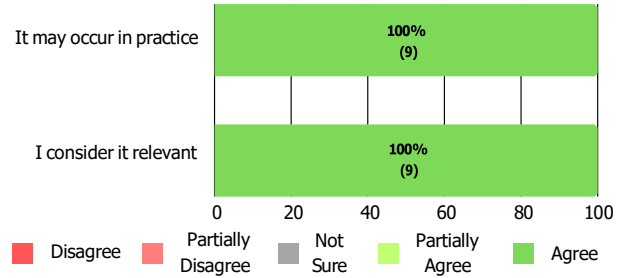

Fig. 5: Inappropriate/Wrong data integration

Participant P6 provided an example from project context: *"In my team, we use a template that streamlines data integration. Before adopting this approach, we encountered challenges as multiple team members were tackling the same issue independently, resulting in redundant efforts"*.

Participant P5 stated: *"When data integration is done incompletely or inadequately, considering that the machine learning model will learn from this input data, it may not learn effectively, leading to poor accuracy. However, the impact can be even worse, as the model can learn incorrectly if data integration is done incorrectly. An improper shortcut when integrating data can lead to significantly more effort in the future."*.

**Activity: Data consumption**. This activity comprises consuming data from sources.

1) *Missing data consumption:*

Figure 6 depicts the overall votes from the experts and confirms that this issue does not fall under the category of TD and can be discarded.

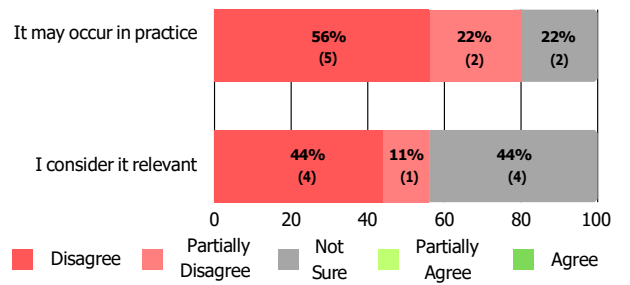

Fig. 6: Missing data consumption

Participant P1 expressed skepticism about the feasibility of continuing development in this scenario, stating: *"I do not see the possibility of continuing development in* this case" while Participant P2 asserted that *"this is a crucial step missing; it would be a bug"*.

2) ***Incomplete/Insufficient data consumption:***
Figure 7 illustrates the results from both sessions. Most participants reported agreement, leading us to conclude that this issue occurs and can be categorized as highly relevant.

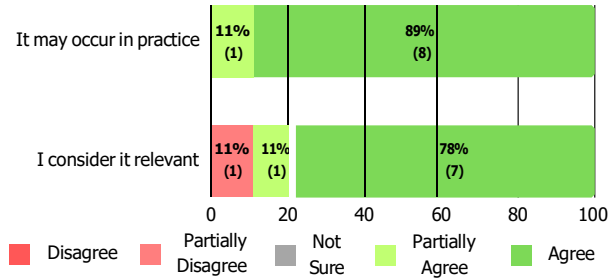

Fig. 7: Incomplete/Insufficient data consumption

Participant P3 explained that *"This is a challenge we often encounter, and we deal with it by filling in the gaps by repeating the previous value, typically the standard procedure, or by applying appropriate calculations for the situation. It's more a matter of recognizing the situation and establishing a policy to address it. A weak policy to speed delivery without providing an optimal solution may lead to technical debt."*.

Participant P5 stated: *"For me, this problem is very similar to incomplete data integration, but there it involves columns, and here it relates to rows. If, by chance, the most relevant lines are the ones I did not consume, this could become a problem that would be hard to fix later."*.

3) ***Inappropriate/Wrong data consumption:***
The results are shown in Figure 8. Since nearly all participants agreed with both statements, we can confidently conclude that this issue can be classified as highly relevant.

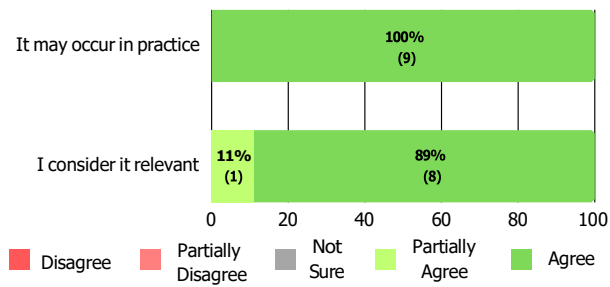

Fig. 8: Inappropriate/Wrong data consumption

Participant P1 made an important statement about the consequences: *"I tend to believe that when something is done incorrectly, it's much more serious for me than when it's not done at all, and I even prefer not to do it than to do it wrong"*.

Participant P8 cited a previous experience: *"This has happened to me. I used a different version of the dataset to compare it with a model generated from another dataset. In theory, we wanted to compare models built with the same dataset. When we evaluated the performance, it was very poor. It wasn't immediately obvious because the model was working."*.

A similar analysis for all other ML workflow phase issues can be found in our online repository [3]. A synthesis of the overall results follows.

*D. Results*

Table III provides a list of issues that, according to the focus group discussions, may occur leading to ML code TD, grouped by ML workflow phase. These issues appear ordered by their relevance, according to the ML experts, and in alphabetic order for issues with similar relevance.

In the following, we present the results for each of the research questions posed.

*1) RQ1. What are potential issues that could lead to ML code TD?:* To address this question, Table I outlines fourteen key activities spanning the ML system workflow phases. These activities are typical tasks developers perform routinely throughout the ML workflow. Each activity was associated with specific problem types—Missing, Incomplete/Insufficient, or Inappropriate/Wrong [8]. This mapping resulted in 34 identified candidate *issues*. Some specific issues, such as *Incomplete/Insufficient rebalancing*, were excluded from the final list due to their lack of practicality.

Notably, no ML code-related candidate issues were identified in the Problem Understanding and Requirements, Model Deployment, and Model Monitoring phases, which are typically not directly related to producing ML code.

*2) RQ2. To what degree do the identified issues manifest in practice, contributing to code-related TD?:* To determine whether the identified issues occur in practice with the potential of leading to ML code TD, two focus group sessions were conducted with nine experts (four in the first session and five in the second). These sessions evaluated the occurrence of each issue as leading to TD and its relevance.

With respect to occurrence, the discussions refined the initial list of 34 candidate issues (*cf.* Table I) to 30 issues recognized as occurring in practice with the potential of contributing to ML code TD.

*3) RQ3. How do practitioners perceive the relevance of the issues in terms of leading to TD?:* Overall, 24 of the 30 issues that occur in practice and may lead to TD were perceived as highly relevant. Figure 9 illustrates the number of issues per ML workflow phase and summarizes these numbers.

It is possible to observe that, during the data collection phase, four of the five issues that can lead to code-related TD in the focus group sessions were considered highly relevant, and one was considered of low relevance. In the data pre-processing phase, 14 of the 17 identified issues were considered highly relevant, while the remaining three were considered of low relevance. In the model creation and training phase, the seven issues were split into five highly relevant ones and two of low relevance. Lastly, the only issue identified in the model evaluation phase was considered highly relevant.

TABLE III: Issues that lead to ML code TD and their perceived relevance

| Phase | Issue | Relevance |
|---|---|---|
| Data Collection | **Inappropriate/Wrong** data consumption | High |
| | **Inappropriate/Wrong** data integration | High |
| | **Incomplete/Insufficient** data consumption | High |
| | **Incomplete/Insufficient** data integration | High |
| | **Missing** data integration | Low |
| Data pre-processing | **Inappropriate/Wrong** feature selection | High |
| | **Inappropriate/Wrong** missing value identification | High |
| | **Inappropriate/Wrong** outlier identification | High |
| | **Inappropriate/Wrong** pre-processing scaling | High |
| | **Inappropriate/Wrong** rebalancing | High |
| | **Inappropriate/Wrong** removal of inconsistent data | High |
| | **Incomplete/Insufficient** missing value identification | High |
| | **Incomplete/Insufficient** pre-processing scaling | High |
| | **Incomplete/Insufficient** removal of inconsistent data | High |
| | **Missing** features selection | High |
| | **Missing** missing value identification | High |
| | **Missing** outlier identification | High |
| | **Missing** pre-processing scaling | High |
| | **Missing** removal of inconsistent data | High |
| | **Incomplete/Insufficient** feature selection | Low |
| | **Incomplete/Insufficient** outlier identification | Low |
| | **Missing** rebalancing | Low |
| Model creation and training | **Inappropriate/Wrong** hyperparameter tuning | High |
| | **Inappropriate/Wrong** splitting of training/test/validation data | High |
| | **Missing** hyperparameter tuning | High |
| | **Missing** splitting training/test/validation data | High |
| | **Missing** testing candidate algorithm possibilities | High |
| | **Incomplete/Insufficient** hyperparameter tuning | Low |
| | **Incomplete/Insufficient** testing candidate algorithm possibilities | Low |
| Model evaluation | **Incomplete/Insufficient** evaluation metric selection | High |

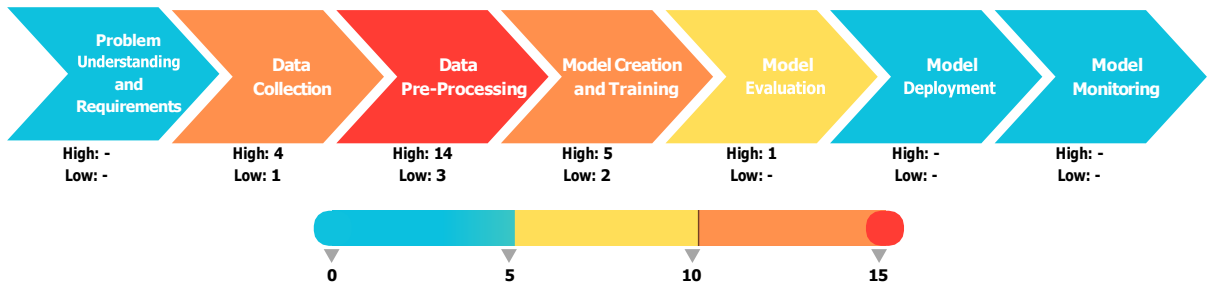

Fig. 9: Heat map of numbers of issues leading to ML code TD per ML workflow phase

The findings emphasize the criticality of addressing ML code TD. The discussions with the participants made it clear that delivering suboptimal solutions that reflect these identified issues can increase the effort required to maintain and evolve ML systems. Furthermore, participants explicitly asked for the results of the final compilation of issues as attention points to avoid accruing technical debt in ML code.

## VII. Discussion

This study highlights issues in different phases of the ML workflow that can contribute to code-related TD. In the following, we discuss the main issues for each ML workflow phase (*cf.* Table III) and their connection with ML code TD, based on the understanding gathered from the focus groups with the experts. We also briefly discuss potential solution options for mitigating these issues.

### A. Data Collection Phase

Issues of the data collection phase related to inappropriate or incomplete data consumption and integration were considered significant contributors to technical debt (TD) in ML code. Inappropriate or incomplete handling of data can embed inefficiencies and workarounds into the codebase, increasing complexity and reducing maintainability. These issues lead to incomplete or biased datasets, resulting in unreliable models and requiring costly rework, such as retraining and pipeline adjustments.

As solution options to mitigate these issues, ML practitioners should focus on implementing robust data validation, scalable integration pipelines, and comprehensive data utilization strategies early in the development process.

### B. Data Pre-Processing Phase

The data pre-processing phase is critical for ensuring the quality of input data in ML systems, and deficiencies at this phase were found to significantly contribute to TD in ML code. Highly relevant issues, such as inappropriate feature selection, missing value identification, and outlier detection, lead to suboptimal models by introducing bias or reducing generalizability. These issues propagate problems throughout the ML pipeline, requiring substantial future effort to correct and retrain models.

Issues in pre-processing tasks like scaling, rebalancing, and handling inconsistencies can further contribute to TD. Inappropriate or incomplete pre-processing introduces inconsistencies in data distributions, leading to poorly performing models that require retraining. For example, incorrect rebalancing or inappropriate removal of inconsistent data forces developers to revisit and refine the logic for handling data, increasing both development time and maintenance costs. Such problems often result in "patch fixes" rather than sustainable solutions, further leading to code TD.

As a solution option to mitigate these issues, practitioners could prioritize automated and rigorous pre-processing pipelines that ensure accuracy and completeness, while integrating quality assurance checks at each step.

### C. Model Creation and Training Phase

The model creation and training phase is central to the development of robust ML systems, and some issues at this phase were also found to significantly contribute to TD in ML code. In particular, inappropriate (or missing) splitting of training, testing, and validation datasets leads to models that are poorly evaluated or fail to generalize effectively. These issues embed suboptimal validation processes into the codebase, requiring extensive future rework to improve model performance and ensure reliable predictions.

The lack of rigorous testing of candidate algorithms further contributes to TD by embedding potentially suboptimal algorithms into the pipeline. Without sufficient testing, developers may employ models that underperform, requiring later replacement or enhancements.

Similarly, inappropriate (or missing) hyperparameter tuning leads to TD by creating inefficiencies in model optimization. Poorly tuned models often perform suboptimally, necessitating repeated adjustments and retraining cycles. This not only increases development time but also introduces complexity into the codebase as developers implement temporary fixes or heuristics to manage the immediate effects of poorly tuned parameters.

As solution options to mitigate these issues, ML practitioners should prioritize systematic dataset splitting, thorough testing of algorithms during model creation, and automated hyperparameter tuning.

### D. Model Evaluation Phase

Incomplete selection of evaluation metrics was considered as an issue of high relevance contributing to TD in the ML code. Indeed, when evaluation metrics fail to align with project objectives or real-world requirements, they can produce misleading assessments of model performance. This often leads to the selection or deployment of models that appear effective under the chosen metric but perform poorly in practical applications. Addressing such misalignments later typically requires significant re-engineering of evaluation logic, retraining of models, and, in some cases, revisiting earlier stages of the ML pipeline. These efforts embed additional complexity into the codebase, increasing maintenance costs, and reducing long-term system reliability.

To mitigate this issue, ML practitioners should emphasize rigorous metric selection that reflects both the technical and business goals of the project. Employing domain expertise, engaging stakeholders, and conducting sensitivity analyses on metric choices can reduce the risk of incurring TD while ensuring that models deliver meaningful and reliable results.

### E. Implications and Future Directions

The findings emphasize the importance of addressing issues that may lead to ML code TD early in the ML workflow.

Similar to what is typically recommended in conventional software engineering, attention to earlier phases is required to engineer robust ML systems. We believe that sharing the compilation of issues that may lead to ML code TD can help to raise awareness of particular attention points to avoid accruing technical debt. Furthermore, the list can provide a foundation for further research into tailored ML code TD mitigation strategies.

## VIII. Limitations and Threats to Validity

We discuss the limitations and threats to validity of our study based on the categories presented by Runeson *et al.* [18].

### A. Internal Validity

A potential threat to internal validity lies in the influence of group dynamics during the focus group discussions. Participants may have been swayed by dominant opinions or the majority consensus, potentially leading to biased assessments of the issues. Furthermore, the discussions of some issues focused more on data handling and model accuracy than on code maintainability. This misalignment with the study's definition of code debt may have influenced the analyses. To mitigate this, the sessions were moderated by two experienced researchers who reinforced the scope of the study when needed, and participants were encouraged to provide independent responses before engaging in group discussions.

### B. Construct Validity

Construct validity may be threatened by ambiguities in the definition of TD and the framing of the issues presented during the sessions. If participants interpreted TD differently or misunderstood specific issues, their evaluations could deviate from the intended research objectives. To address this, a clear and consistent definition of TD and detailed explanations of the issues were provided during the sessions, along with opportunities for participants to ask clarifying questions. Regarding the issues, a limitation of our study is the possibility of unintentionally omitting critical ML code-related issues not addressed in our initial list (Table I). It is noteworthy that we tried to compile our initial list systematically based on typical activities conducted during the ML workflow [2] and typical problem types [8]. We observed that participants mentioned some additional issues during both focus group sessions, *e.g.*, model deployment and consumption, which were, however, not directly related to the ML code. The participants' interest in the final list also increased our confidence in having produced a meaningful initial list to subsidize the discussions.

### C. External Validity

The external validity of the findings may be limited by the number of participants and the number of focus group sessions. Notably, it is suggested to plan online focus groups with fewer participants than face-to-face focus groups [14], with four considered appropriate [13]. Still, we know that more than two focus group sessions are needed to reach generalizable findings. Unfortunately, identifying ML experts with experience and willing to collaborate voluntarily is not a trivial task. However, we believe that the two focus group sessions already allowed us to improve our understanding and gather some valuable insights to be shared.

### D. Reliability

To enhance reliability, the study followed a structured methodology, including standardized discussion templates and consistent facilitation across sessions, according to the guidelines by Kontio *et al.* [11] and additional advice for remote focus groups by *Menary et al.* [15]. Following the advice by Runeson *et al.* [19], we made our main artifacts available online [3], including the Miro board used to subsidize the discussions (containing all the votes and post-it comments), and the complete analysis conducted to refine the list of issues.

## IX. Conclusion

This study investigated issues across the ML workflow that contribute to ML code TD. As part of this effort, we identified 34 potential issues based on typical ML development activities and problem types. These were refined to 30 issues through two focus group sessions with nine experienced ML practitioners. During these sessions, each issue was assessed for its occurrence in practice as leading to TD and its perceived relevance, resulting in 24 issues being classified as highly relevant. The data pre-processing phase emerged as the most critical, with 14 highly relevant issues, followed by data collection, model creation and training, and model evaluation.

The focus groups provided practitioner insights into real-world challenges, shedding light on issues that contribute to TD in ML code and on their consequences, such as degraded model performance and increased maintenance effort over time. We believe that making our findings and the list of issues publicly available can help practitioners to address these issues early, minimizing TD accumulation and enhancing the long-term maintainability of ML-related code.

While this study focuses exclusively on ML code-related TD, we acknowledge the existence of other forms of TD that are also relevant to ML systems, such as data or architectural TD. Exploring these additional forms could provide a more comprehensive understanding of TD in ML systems. Furthermore, our investigations of the issues leading to ML code-related TD involved conducting two focus group sessions. Future research could employ complementary empirical strategies, such as longitudinal case studies, to refine the list of issues and strengthen our findings.

## Acknowledgment

The authors sincerely thank the participants of the focus group sessions for their time and invaluable contributions, which greatly enriched this research. We also express our gratitude to FAPERJ (Rio de Janeiro State Research Support Foundation) for grant E-26/204.256/2024, CNPq (National Council for Scientific and Technological Development, Brazil) for grant 312275/2023-4, and Stone Co. (project 1006) for their generous support.